\documentclass[conference]{IEEEtran}

\usepackage{color}
\usepackage{graphicx}
\usepackage{amsmath}
\usepackage{epstopdf}
\usepackage{amsthm}
\usepackage{amssymb}
\usepackage{epsfig}
\usepackage{amsfonts}
\usepackage[]{algorithmic}
\usepackage[]{algorithm2e}
\usepackage{listings}
\usepackage[keeplastbox]{flushend}
\usepackage{fancybox}
\usepackage{epsfig}
\usepackage{mathtools}
\usepackage[]{algorithm2e}
\DeclareRobustCommand{\rchi}{{\mathpalette\irchi\relax}}
\newcommand{\irchi}[2]{\raisebox{\depth}{$#1\chi$}} 
\usepackage{amsmath}
\ifCLASSINFOpdf
\else
\fi

\begin{document}
\title{A compressive channel estimation technique robust to synchronization impairments}

\author{\IEEEauthorblockN{Nitin Jonathan Myers and Robert W. Heath Jr.}
\IEEEauthorblockA{Department of Electrical and Computer Engineering\\
The University of Texas at Austin\\
Email: $\{$nitinjmyers, rheath$\}$@utexas.edu
}
}

\maketitle

\begin{abstract}
Initial access at millimeter wave frequencies is a challenging problem due to hardware non-idealities and low SNR measurements prior to beamforming. Prior work has exploited the observation that mmWave MIMO channels are sparse in the spatial angle domain and has used compressed sensing based algorithms for channel estimation. Most of them, however, ignore hardware impairments like carrier frequency offset and phase noise, and fail to perform well when such impairments are considered. In this paper, we develop a compressive channel estimation algorithm for narrowband mmWave systems, which is robust to such non idealities. We address this problem by constructing a tensor that models both the mmWave channel and CFO, and estimate the tensor while still exploiting the sparsity of the mmWave channel. Simulation results show that under the same settings, our method performs better than comparable algorithms that are robust to phase errors.
\end{abstract}
\IEEEpeerreviewmaketitle

\section{Introduction}
Millimeter wave (mmWave) communication is a potential candidate for 5G systems due to the enormous amount of spectrum available at mmWave frequencies \cite{ranganmmwave}. Such systems are likely to employ large antenna arrays and use highly directional beamforming to provide sufficient received signal power \cite{heathoverview}. Channel estimation at mmWave, however, is challenging as the best beams have to be estimated using low SNR measurements \cite{heathoverview}. A naive way to estimate the best beams would be to perform an exhaustive beam search at the transmitter (TX) and receiver (RX) to operate with the beam pair corresponding to the maximum received SNR. This method, however, incurs a lot of training overhead.
\par
Unlike the lower frequency systems, MIMO channels at mmWave are marginally sparse in the spatial angle domain due to clustering in the propagation environment \cite{heathoverview}. The sparse nature of mmWave MIMO channels has been exploited to estimate the channel using compressed sensing (CS) \cite{ramacs}\cite{alkhcs}\cite{agile}. Most of the existing methods, however, assume an ideal CS measurement model and fail to perform well in the presence of hardware impairments like phase noise, carrier frequency offset (CFO) etc. Although the exhaustive beam search method (which considers just the magnitude of beam space measurements) is robust to such non idealities, it does not make use of the fact that mmWave channels are sparse. Hence, there is a need to develop better channel estimation algorithms that are simultaneously robust to such non-idealities and require fewer measurements by exploiting the sparsity of mmWave channel. To the best of our knowledge, the only work that addresses this problem is \cite{agile}, where the optimal beam pair for analog beamforming is estimated using the components of the channel matrix along specially designed measurement matrices, that satisfy hardware constraints. Although it performs better than the conventional CS based algorithms that do not model CFO, it ignores noise and neglects the phase of measurements in its model. Our approach handles both these issues.
\par In this paper, we propose a compressive algorithm to jointly estimate the CFO and channel for narrowband analog beamforming systems (having a single RF chain) with uniform linear arrays (ULAs) at the base station (BS) and the mobile station (MS). Extension of our method to uniform planar arrays is straightforward. We assume that timing synchronization is already performed and that the channel is sparse per the virtual channel model \cite{heathoverview}. Our key contributions are in modeling the MIMO channel and CFO using a third order tensor and compressively estimating the tensor using the available measurements. We highlight the fact that unlike in \cite{agile} where just the angle-of-arrivals (AoAs) and angle-of-departures (AoDs) are found, our method estimates the channel matrix and CFO. Further, simulations show that our algorithm performs better than that proposed in \cite{agile} for the same setting.
\par We use the following notation: $\mathbf{A}$ is a matrix, $\mathbf{a}$ is a column vector, $\mathcal{A}$ is a tensor and $a, A$ denote scalars. Using this notation, $\bar{a}$ is the complex conjugate of $a$, $\mathbf{A}^{\ast}$ is the conjugate transpose of $\mathbf{A}$ and $\mathbf{A}^{(i)}$ denotes the $i^{th}$ row of $\mathbf{A}$. We use $\left[N\right]$ to denote the set $\left\{ 1,2,3,..N\right\}$. The symbols $\circledcirc$ and $\otimes$ are used to denote the outer product \cite{siritensors} and kroenecker product respectively. The matrix $\mathbf{U}_N \in \mathbb{C}^{N \times N}$ denotes a  DFT matrix of dimension $N$ and is given by $\mathbf{U}_N\left(k,\ell\right)=e^{-j\frac{2\pi (k-1)(\ell-1)}{N}}$, for $k,\ell \in \left[N\right]$. We define $\mathbf{e}_n \in \mathbb{C}^{M\times 1}$ to be a cannonical basis vector with its $n^{th}$ entry as 1.

\section{Preliminaries on Tensors}
In this section, we provide necessary definitions from tensor algebra that will aid in  understanding the subsequent sections. We limit our discussion to tensors of order 3, as it suffices to model our problem.
\par A tensor is a multidimensional array, which is essentially an extension of vectors and matrices. For example, a matrix $\mathbf{X} \in \mathbb{C} ^{N_1 \times N_2}$ is a tensor of order 2 and dimension $N_1N_2$. Similarly $\mathcal{A} \in \mathbb{C} ^{N_1 \times N_2 \times N_3}$ is a tensor of order 3 and has dimension $N_1N_2N_3$. For two tensors $\mathcal{A},\mathcal{B}$, their inner product is defined as 
\begin{equation}
\left\langle \mathcal{A},\mathcal{B}\right\rangle =\sum_{{k\in[N_{3}]},{j\in[N_{2}]},{i\in[N_{1}]}}\mathcal{A}\left(i,j,k\right)\bar{\mathcal{B}}\left(i,j,k\right), 
\end{equation} and the norm of a $\mathcal{A}$ is given by $\sqrt{\left\langle \mathcal{A},\mathcal{A}\right\rangle }$. The $\ell_1$- norm of $\mathcal{A}$ is given by
\begin{equation}
\left\Vert \mathcal{A}\right\Vert_{\ell_1}=\sum_{i=1}^{N_{1}}\sum_{j=1}^{N_{2}}\sum_{k=1}^{N_{3}}\left|\mathcal{A}\left(i,j,k\right)\right|.
\end{equation}
The mode $3$ unfolding of $\mathcal{A}$ is a matrix denoted by $\mathcal{A}_{(3)} \in \mathbb{C}^{N_3 \times N_1N_2}$  and is given by $\mathcal{A}_{(3)}=$
\begin{equation}
\left[\mathrm{vec}\left(\mathcal{A}\left(:,:,1\right)\right),\mathrm{vec}\left(\mathcal{A}\left(:,:,2\right)\right),..,\mathrm{vec}\left(\mathcal{A}\left(:,:,N_{3}\right)\right)\right]^T.
\end{equation}

\section{System model}
\label{sec:system model}
In this paper, we focus on initial access in narrowband mmWave systems with analog beamforming at the TX and RX. Consider a point-to-point link with ULAs of $N_t$ and $N_r$ antennas at the BS (TX) and MS (RX) respectively. The antenna arrays at the TX and RX are connected to their corresponding RF chain through a network of digitally controlled phase shifters. For the $n^{\mathrm{th}}$ measurement, let $\mathbf{f}_{n}$ and $\mathbf{w}_{n}$ be the unit norm beam training vectors applied to the phase shifters at the TX and RX respectively, with $\sqrt{N_t}\mathbf{f}_{n}\in\mathbb{Q}^{N_{t}}$, $\sqrt{N_r}\mathbf{w}_{n}\in\mathbb{Q}^{N_{r}}$. As the phase shifters are digitally controlled, we have $\mathbb{Q}=\left\{ e^{j\theta_{1}},e^{j\theta_{2}},...,e^{j\theta_{q}}\right\} $, where $\theta_{i}=\frac{2\pi i}{q}$ for a $q$ level phase quantization. For a sequence of $M$ measurements, let $\mathbf{r}\left[n\right]$ be the received symbol and $\left(\mathbf{y}\left[n\right]=\frac{\mathbf{s}^{\ast}\left[n\right]\mathbf{r}\left[n\right]}{\left\Vert \mathbf{s}\left[n\right]\right\Vert^2 }\right)$ be the measurement corresponding to the transmitted symbol $\mathbf{s}\left[n\right]$, with $\left\Vert \mathbf{s}{\left[n\right]}\right\Vert =\sqrt{\rho},\,\forall n \in \left[ M\right]$. The $n^{th}$ received symbol and measurement are given by 
\begin{align}
\nonumber
\mathbf{r}\left[n\right]&=\mathbf{w}_{n}^{\ast}\mathbf{H}\mathbf{f}_{n}e^{j\left(\omega_{e}n+\phi_{n}\right)}\mathbf{s}{\left[n\right]}+\tilde{\mathbf{v}}{\left[n\right]} \,\,\,\,\,\, \\
\label{eq:model}
\mathbf{y}{\left[n\right]}&=\mathbf{w}_{n}^{\ast}\mathbf{H}\mathbf{f}_{n}e^{j\left(\omega_{e}n+\phi_{n}\right)}+\mathbf{v}{\left[n\right]}, &&\forall n\in \left[M\right]
\end{align}
where $\mathbf{H}\in  \mathbb{C}^{N_{r}\times N_{t}}$ is the channel matrix; $\tilde{\mathbf{v}} \sim \mathcal{CN}\left(0,\sigma^2\mathbf{I}_{M\times M}\right)$, $\mathbf{v} \sim \mathcal{CN}\left(0,\frac{\sigma^2}{\rho}\mathbf{I}_{M\times M}\right)$; $\phi_{n|\phi_{n-1}} \sim \mathcal{N}\left(\phi_{n-1},\tau^{2} \right)$ is considered to be a Wiener phase noise process \cite{oscill} (with $\phi_0=0$); $\omega_e=2 \pi f_e T $, with $T$ as the symbol duration and $f_e$ (Hz) as the carrier frequency offset.
\par Consider a propagation environment with $N_{\mathrm{c\ell}}$ clusters and $N^n_{\mathrm{ray}}$ rays in the $n^{\mathrm{th}}$ cluster. For the $m^{\mathrm{th}}$ path of the  $n^{\mathrm{th}}$ cluster, let $\gamma_{n,m}$ denote the complex gain and $\theta^r_{n,m}\left(\theta^t_{n,m}\right)$ denote the AoA(AoD). Let $\lambda$ be the carrier wavelength and $d$ be the antenna spacing in the ULAs at the BS and MS. With $\omega_{n,m}^{\mathrm{r}} \coloneqq \frac{2 \pi d}{\lambda}\sin(\theta_{n,m}^{\mathrm{r}})$, $\omega_{n,m}^{\mathrm{t}} \coloneqq \frac{2 \pi d}{\lambda}\sin(\theta_{n,m}^{\mathrm{t}})$ and the following definition 
\begin{equation}
\label{eq:arrresp}
\mathbf{a}_{_{N}}\left(\theta\right)=\left[1\,e^{j\theta}\,e^{j2\theta}\,\cdots\,e^{j(N-1)\theta}\right]^{T},
\end{equation}
the MIMO channel matrix $\mathbf{H}$, in baseband is given by 
\begin{equation}
\label{eq:mmwave}
\mathbf{H}=\frac{1}{\sqrt{N_{\mathrm{c\ell}}}}\sum_{n=1}^{N_{\mathrm{c\ell}}}\frac{1}{\sqrt{N_{\mathrm{ray}}^{n}}}\sum_{m=1}^{N_{\mathrm{ray}}^{n}}\gamma_{n,m}\mathbf{a}_{_{N_{r}}}\left(\omega_{n,m}^{\mathrm{r}}\right)\mathbf{a}_{_{N_{t}}}^{\ast}\left(\omega_{n,m}^{\mathrm{t}}\right).
\end{equation}
\par At mmWave carrier frequencies, $\mathbf{H}$ in (\ref{eq:mmwave}) is marginally sparse matrix in the spatial DFT basis \cite{heathoverview}. The channel matrix would be exactly sparse if the spatial frequencies align exactly on the grid; when it is not the case, a grid of finer resolution can be chosen to increase compressibility at the expense of higher dimensionality. We assume that the spatial frequencies (of the form $\left(\omega_{x},\omega_{y}\right)$ ) of $\mathbf{H}$ come from a discrete set, i.e, $\omega_{x}\in\left\{ 0,\frac{2\pi}{N_{t}},\frac{4\pi}{N_{t}},..,\frac{2\pi\left(N_{t}-1\right)}{N_{t}}\right\} ,\omega_{y}\in\left\{ 0,\frac{2\pi}{N_{r}},\frac{4\pi}{N_{r}},..,\frac{2\pi\left(N_{r}-1\right)}{N_{r}}\right\}$ for our analysis; off grid extensions can be made using  \cite{ramacs}\cite{duartescs}. We test our algorithm using off-grid parameters in the simulations.  The channel matrix in (\ref{eq:model}) can be expressed as
\begin{equation}
\label{eq:Htensor}
\mathbf{H}=\sum_{i=1}^{N_{r}}\sum_{j=1}^{N_{t}}\alpha_{_{ij}}\mathbf{a}_{_{N_{r}}}\left(\frac{2\pi i}{N_{r}}\right)\circledcirc\mathbf{a}_{_{N_{t}}}\left(\frac{2\pi j}{N_{t}}\right)\mathrm{\,,}
\end{equation} 
With the assumption that $\mathbf{H}$ is sparse in the spatial DFT basis, $\mathbf{C} \coloneqq \left\{ \mathbf{\alpha}_{_{ij}}\right\} _{i,j=1}^{N_{r},N_{t}}$ is now a sparse matrix, of sparsity $K$ in $N_rN_t$ dimension. 
\section{Modeling CFO and the channel using tensors}
The notion behind modeling the channel and CFO in a tensor comes from the intuition to consider the spatial frequencies corresponding to AoAs, AoDs and the CFO ($\omega_e$) along three different dimensions. It may be observed from (\ref{eq:model}) that $\mathbf{y}\left[n\right]$ is a noisy version of the inner product between an unknown matrix ($\mathbf{H}e^{j\Omega_n}$) with a known measurement matrix $\mathbf{w}_n\mathbf{f}^{\ast}_n$, where $\Omega_n=\omega_e n+\phi_n$. We model a collection of $M$ such matrices using a tensor $\rchi \in \mathbb{C}^{N_{r}\times N_{t}\times M}$ such that the $k^{th}$ frontal slab  \cite{siritensors} of $\rchi$ is given by $\rchi(:,:,k)=\mathbf{H}e^{j\Omega_k}$. Hence 
\begin{equation}
\label{eq:tensor}
\rchi= \mathbf{H} \circledcirc \mathbf{e}_{_{\Omega}} \mathrm{\,,}
\end{equation}  
where $\mathbf{e}_{_{\Omega}}=\left(e^{j\Omega_{1}},e^{j\Omega_{2}},..,e^{j\Omega_{M}}\right)^T$. We expand $\mathbf{e}_{_{\Omega}}\in \mathbb{C}^{M\times1}$ in the $M$ dimensional discrete fourier basis as 
\begin{equation}
\label{eq:e_omega}
\mathbf{e}_{_{\Omega}}=\sum_{k=1}^{M}\beta_{_k}\mathbf{a}_{_M}\left(\frac{2\pi k}{M}\right),
\end{equation}
with $\mathbf{a}_{_M}\left(.\right)$ defined according to (\ref{eq:arrresp}). Unlike $\mathbf{C}$, $\mathbf{p}\triangleq \left(\beta_{1},\beta_{2},..,\beta_{M}\right)^T$ is not assumed to be exactly sparse because it is unrealistic to assume that CFO ($\omega_e$) lies on the grid. Moreover phase noise distorts the spectrum corresponding to CFO. From (\ref{eq:Htensor}), (\ref{eq:tensor}) and (\ref{eq:e_omega}), the tensor $\rchi$ can be expressed as
\begin{equation}
\label{eq:chitensor}
\rchi=\sum_{{i,j,k}}\,\mathcal{G}\left(i,j,k\right)\mathbf{a}_{_{N_{r}}}\left(\frac{2\pi i}{N_{r}}\right)\circledcirc\mathbf{a}_{_{N_{t}}}\left(\frac{2\pi j}{N_{t}}\right)\circledcirc\mathbf{a}_{_M}\left(\frac{2\pi k}{M}\right),
\end{equation} 
where $\mathcal{G} \in \mathbb{C}^{N_{r}\times N_{t}\times M}$ is another tensor, with $\mathcal{G}\left(i,j,k\right)=\alpha_{_{ij}}\beta_{_k}$. Using the tensor notation, the measurements in (\ref{eq:model}) can now be expressed as 
\begin{align}
\nonumber
\mathbf{y}{\left[n\right]}&=\left\langle \rchi,\,\mathbf{w}_{n}\circledcirc\bar{\mathbf{f}}_{n}\circledcirc\mathbf{e}_{n}\right\rangle +\mathbf{v}\left[n\right] \\
\label{eq:tensormeas}
 &=\left\langle \mathcal{G},\mathcal{M}_n\right\rangle +\mathbf{v}\left[n\right], 
\end{align}
where $\mathcal{M}_n=\mathbf{\,U}_{N_{r}}\mathbf{w}_{n}\circledcirc\mathbf{U}_{N_{t}}\bar{\mathbf{f}}_{n}\circledcirc\mathbf{U}_{M}\mathbf{e}_{n}$. We have thus modelled the channel and CFO using a tensor $\rchi$, and their joint estimation is now a tensor estimation problem.
\begin{figure}[h!]
\begin{center}
\includegraphics[width=1.01 \linewidth]{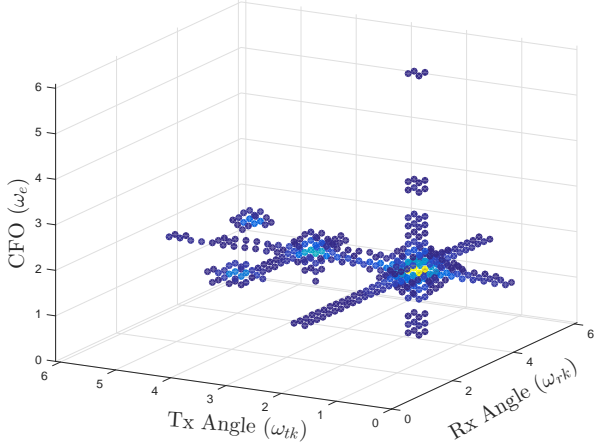}
  \caption{Magnitude plot of the third order tensor $\mathcal{G}$, for $N_r=N_t=32$, $N_{\mathrm{c\ell}}=4$ clusters , CFO $\omega_e =\frac{2\pi}{3}, \tau=0.1$. For illustrative purposes as well as simulations, the on grid assumption on $\omega_{rk},\omega_{tk}$ is waived.}
  \label{fig:tensor}
\end{center}
\end{figure}
\par Although $\rchi$ is a tensor of size $N_rN_tM$, it is just an outer product of a rank $K$ matrix with a vector and hence $\mathrm{tensor\,rank}\left(\rchi\right) \leq K$
\cite{siritensors}. Further, $\mathcal{\rchi}$ is a tensor compressible in the 3D DFT basis, due to sparsity of $\mathbf{C}$ and marginal sparsity of $\mathbf{p}$ (as $\mathcal{G}_{_{ijk}}= \alpha_{_{ij}}\beta_{_k}$), as shown in Fig.\ref{fig:tensor}. Hence, low rank recovery \cite{lowranktensor} or CS based  \cite{tensorcs} algorithms for tensors can be used to reconstruct $\rchi$. We use compressed sensing because the sparsity basis of $\rchi$ is known and CS exploits this fact unlike low rank tensor recovery algorithms.
\section{Compressed sensing based recovery}
Similar to the vector case in which a sparse vector can be recovered using projections onto a carefully chosen lower dimensional subspace, a higher dimensional sparse tensor can also be recovered from a lower dimensional one, under certain constraints. An extensive theory on  CS for sparse tensor recovery can be found in \cite{tensorcs}. The recovery of $\mathcal{G}$ in (\ref{eq:tensormeas}) using the standard $\ell_1$ minimization algorithm can be given as,
\begin{equation}
\label{eq:tensorl1}
\mathrm{minimize\,\,} \left\Vert \mathcal{F}\right\Vert _{\ell_{1}}
\end{equation}
\begin{center}
$\mathrm{s.t\,}\sum_{n=1}^{M}\left|\mathbf{y}\left[n\right]-\left\langle \mathcal{F},\mathcal{M}_{n}\right\rangle \right|^{2}\leq\sigma_{_{N}}^{2}.$
\end{center} 
\par The simplest way to solve (\ref{eq:tensorl1}) would be to use Kroenecker Compressed Sensing           (KCS), i.e, to vectorize the tensor and then apply the well known techniques of vector compressed sensing to this problem. Though KCS can recover the sparse tensor, it has huge complexity and the structural information conveyed by the tensor may be lost due to vectorization. We use orthogonal matching pursuit (OMP), a greedy algorithm for sparse tensor recovery. For a given set of measurement tensors $\left\{ \mathcal{M}_{n}\right\} _{n=1}^{M}$, we define a linear operator $\mathcal{P}\mathrm{:}\,\mathbb{C}^{N_1 \times N_2 \times M} \rightarrow \mathbb{C}^M$ and an adjoint $\mathcal{P}^{\ast}\,\mathrm{:}\mathbb{C}^{M} \rightarrow \mathbb{C}^{N_1 \times N_2 \times M}$ as 
\begin{align}
\label{eq:proj}
\mathcal{P}\left(\mathcal{G}\right)&=\left(\left\langle \mathcal{G},\mathcal{M}_{1}\right\rangle ,\left\langle \mathcal{G},\mathcal{M}_{2}\right\rangle ,...,\left\langle \mathcal{G},\mathcal{M}_{n}\right\rangle \right)^{T},\\
\label{eq:adj}
\mathcal{P}^{\ast}\left(\mathbf{y}\right)&=\sum_{n=1}^{M}\mathbf{y}{\left[n\right]}\mathcal{M}_{n}.
\end{align}
From (\ref{eq:tensormeas}), (\ref{eq:proj}), we have $\mathbf{y}=\mathcal{P}\left(\mathcal{G}\right)+\mathbf{v}$.
\subsection{Tensor estimation algorithm}
Let $\epsilon \coloneqq \mathbb{E}\left[\mathbf{v}^{\ast}\mathbf{v}\right]$ be the stopping threshold and $N_{\mathrm{iter}}$ be the maximum allowed iterations for the OMP based recovery of $\mathcal{G}$.
\begin{algorithm}[h]
\label{algo-OMP}
\textbf{Input:} $\mathbf{y}$, $\mathcal{P}$, $\epsilon$, $\mathcal{F}^{0}=\mathbf{0}$,$\mathcal{D}^{0}=\phi$, $n=1$;\\ 
 \While{$\left\Vert \mathbf{y}-\mathcal{P}\left(\mathcal{F}^{(n-1)}\right)\right\Vert^2 >\epsilon \,\, \mathrm{and}\,\, \ n\leq N_{\mathrm{iter}}$\\}
 {
  $\Upsilon^{(n)}=\mathrm{argmax}\left|\mathcal{P}^{\ast}\left(\mathbf{y}-\mathcal{P}\left(\mathcal{F}^{(n-1)}\right)\right)\right|$\\
  $\mathcal{D}^{(n)}=\Upsilon^{(n)} \cup \mathcal{D}^{(n-1)}$\\
 $\mathcal{F}^{(n)}=\underset{\mathcal{V}:\mathrm{supp\left(\mathcal{V}\right)}=\mathcal{D}^{\left(n\right)}}{\mathrm{arg\,min}}\left\Vert \mathbf{y}-\mathcal{P}\left(\mathcal{V}\right)\right\Vert _{\ell_{2}}$ \\
 }
 \KwResult{$\hat{\mathcal{G}}=\mathcal{F}^{(n)}$. 
}
\textit{\\}
\caption{OMP to recover $\mathcal{G}$}
\end{algorithm}
\par It may be noted that $\mathrm{argmax}\left|\mathcal{A}\right|$ returns a tuple corresponding to the location where the tensor takes its maximum (in magnitude). In a nutshell, the OMP algorithm iteratively estimates $\mathcal{G}$, with $\mathcal{F}^{(n)}$ being the estimate at the $n^{\mathrm{th}}$ iteration. In every iteration, it identifies a tensor element that maximally explains the residue and finds a tensor accordingly, using a least squares estimate. A detailed treatment on OMP can be found in \cite{OMP}. 
\par Having obtained $\hat{\mathcal{G}}$ using Algorithm \ref{algo-OMP}, we now need to split it into a matrix and a vector, corresponding to  $\mathbf{C}$ and $\mathbf{p}$ respectively, as $\mathcal{G}=\mathbf{C} \circledcirc \mathbf{p}$. This decomposition is performed using singular value decomposition (svd) of $\hat{\mathcal{G}}_{(3)}$, the mode 3 unfolding of $\hat{\mathcal{G}}$. In other words, $\hat{\mathbf{p}}$ and $\mathrm{vec}\left(\bar{\hat{\mathbf{C}}}\right)$ are the left and right singular vectors corresponding to the maximum singular value of $\hat{\mathcal{G}}_{(3)}$. The final step of unfolding the tensor along mode 3 followed by the SVD is inspired by \cite{biconvex}. It may be noted that $\hat{\mathbf{p}}$ is an estimate of $\left(\beta_1,\beta_2,.\,.\,.\beta_M\right)$, the DFT of $\mathbf{e}_{_{\Omega}}$ in (\ref{eq:e_omega}). Hence, an estimate of CFO $\left(\omega_e \right)$ can be obtained from $\hat{\mathbf{p}}$, using a kalman filter based approach, that accounts for the Wiener phase noise. Due to space constraints, we omit the discussion on CFO estimation and focus only on beam alignment in the subsequent sections.
\subsection{Analog beamforming}
With $\hat{\mathbf{C}}$, the channel estimate can be given by $\mathbf{H}_{\mathrm{est}}=\mathbf{U}_{N_{r}}^{\ast}\hat{\mathbf{C}}\mathbf{U}_{N_{t}}^{\ast}$, upto a scale factor. For data transmission, the beamforming vectors $\mathbf{f}_{\mathrm{est}} \in \mathbb{Q}^{N_t}$ and $\mathbf{w}_{\mathrm{est}} \in \mathbb{Q}^{N_r}$ have to be chosen such that $\left|\mathbf{w}^{\ast}_{\mathrm{est}}\mathbf{H}_{\mathrm{est}}\mathbf{f}_{\mathrm{est}}\right|$ is maximized. We find $\mathbf{f}_{\mathrm{est}},\mathbf{w}_{\mathrm{est}}$ by performing the SVD of $\mathbf{H}_{\mathrm{est}}$ followed by element-wise phase quantization of the singular vectors, corresponding to the maximum singular value.  
\subsection{Advantages of modeling using tensors}
The most important advantage of modeling the MIMO channel and CFO estimation problem using tensors is that it maintains the structural information, which is otherwise lost when vectorization is used. Modeling with tensors provides scope for spectral compressed sensing \cite{duartescs}, which deals with problem of off the grid CS. Using the techniques in \cite{duartescs}, finer estimates of CFO and spatial frequencies of the channel matrix can be obtained. Further, it can be seen from Fig.\ref{fig:tensor}, that most of the mass in the tensor $\mathcal{G}$ is concentrated along a set of horizontal slabs around $\omega = \omega_e$. This information can be used to apply structured CS algorithms \cite{structured} to either reduce the number of CS measurements or provide better estimates with the same number of measurements.
\subsection{Reducing the dimension of our problem}
In typical systems $f_e \in \left[-f_{\mathrm{max}},f_{\mathrm{max}}\right]$, where $f_{\mathrm{max}}$ depends on the quality of local oscillators used at TX and RX, and is typically in the order of parts per millions (ppms) of the carrier frequency. As a DFT grid of finite resolution ($\frac{2\pi}{M}$) is used to model CFO, a spectral leakage factor of $\gamma \in \left[1,\left(2Tf_{\mathrm{max}} \right)^{-1}\right]$ is considered and the components that lie on the DFT grid within the range $\left[ - \gamma f_{\mathrm{max}}, \gamma f_{\mathrm{max}} \right]$ are recovered. With $P=\left\lceil M\gamma f_{max}T\right\rceil$, the following approximation for $\mathbf{e}_{_{\Omega}}$ in (\ref{eq:e_omega}) is used.
\begin{equation}
\label{eq:new_eomeg}
\mathbf{e}_{_{\Omega}}=\sum_{k=1}^{P}\beta_{k}\mathbf{a}_{_{M}}\left(\frac{2\pi k}{M}\right)+\sum_{k=M-P}^{M}\beta_{k}\mathbf{a}_{_{M}}\left(\frac{2\pi k}{M}\right)
\end{equation}
The limits of $k$ in (\ref{eq:chitensor}) change accordingly and we now need to solve for a tensor of dimension $N_tN_r\left(2P+1\right)$ rather than $N_tN_rM$. It may be noted that this approximation does not hold for high values of phase noise variance ($\tau$), as the magnitude spectrum of $\mathbf{e}_{_{\Omega}}$ would not be concentrated about $\omega_e$ in such case.
\subsection{Analogy with lifting techniques}
Although we started with the intuition to model spatial frequency and CFO along different dimensions, we have essentially moved to a higher dimensional space. Our framework of modeling CFO, channel using tensors is analogous to lifting techniques that exist in the literature   \cite{phaselift} \cite{biconvex}. The mode 3 unfolding of $\mathcal{G}$ can also be derived using lifting \cite{biconvex}, a method that convexifies a nonconvex optimization problem by moving to a higher dimensional space, and is illustrated below.
\par Let the phase error free measurements be given by $\mathbf{z}\left[n\right]=\mathbf{w}_{n}^{\ast}\mathbf{H}\mathbf{f}_{n}+\mathbf{v}\left[n\right]$. From (\ref{eq:Htensor}), it follows that $\mathbf{H}=\mathbf{U}^{\ast}_{_{N_r}}\mathbf{C}\mathbf{U}^{\ast}_{_{N_t}}$ in the spatial DFT basis. Using properties of kroenecker products, it can be shown that $\mathbf{z}\left[n\right]=\left(\mathbf{f}_{n}^{T}\mathbf{U}_{N_{t}}^{\ast}\right)\otimes\left(\mathbf{w}_{n}^{\ast}\mathbf{U}_{N_{r}}^{\ast}\right)\mathrm{vec}\left(\mathbf{C}\right)+\mathbf{v}\left[n\right]$. On stacking all the phase error free measurements we get $\mathbf{z}=\mathbf{Ax}+\mathbf{v}$, where $\mathbf{A}^{(n)}=\left(\mathbf{f}_{n}^{T}\mathbf{U}_{N_{t}}^{\ast}\right)\otimes\left(\mathbf{w}_{n}^{\ast}\mathbf{U}_{N_{r}}^{\ast}\right)$, and $\mathbf{x}=\mathrm{vec}\left(\mathbf{C}\right)$ is a $K$ sparse vector. The observed measurements in (\ref{eq:model}) can now be given by,
\begin{align}
\nonumber
\mathbf{y}&=\mathrm{diag}\left(e^{j\Omega_{1}},e^{j\Omega_{2}},..,e^{j\Omega_{n}}\right)\mathbf{Ax}+\mathbf{v}\\
\label{eq:biconv1}
&=\mathrm{diag}\left(\mathbf{U}^{\ast}_M\mathbf{p}\right)\mathbf{Ax}+\mathbf{v}
\end{align}
Hence, we have 
\begin{equation}
\label{eq:biconv2}
\mathbf{y}\left[n\right]=\mathbf{U}_{M}^{\ast\left(n\right)}\mathbf{p}\mathbf{A}^{(n)}\mathbf{x}+\mathbf{v}\left[n\right]=\mathbf{U}_{M}^{\ast\left(n\right)}\mathbf{p}\mathbf{x}^{T}\left(\mathbf{A}^{(n)}\right)^{T}+\mathbf{v}\left[n\right]
\end{equation}
By defining $\mathbf{X}=\mathbf{p}\mathbf{x}^T$, we have $\mathbf{y}=\mathcal{P}_1\left(\mathbf{X}\right)+\mathbf{v}$, where $\mathcal{P}_1$ is a linear operator defined in accordance with (\ref{eq:biconv2}). It is proposed in \cite{biconvex} to recover $\mathbf{X}$, by minimizing it's $\ell_1$ norm and then perform SVD to obtain estimates of $\mathbf{p},\mathbf{x}$, upto a scale factor. From a signal processing perspective, the tensor based approach can model and recover a sparse tensor with arbitrary locations of sparsity (eg:- finding three dimensional frequency of a third order tensor). Lifting can be considered as a specific instance of the tensor approach, where the locations of sparsity are constrained around a plane.
\section{Simulation Results}
In this section, we compare the performance of our tensor based algorithm with Agile Link \cite{agile} and OMP (CFO ignored). We consider the system model in Section \ref{sec:system model}, with ULAs of size $N_t=32$ and $N_r=16$, antenna spacing of $d=\frac{\lambda}{2}$ for each of the ULAs and a narrowband mmWave channel in (\ref{eq:mmwave}) with $N_{\mathrm{c\ell}}=2$ clusters, each comprising of $N_{\mathrm{ray}}=10$ paths and $3$ degrees of angular spread. We consider the complex path gains $\gamma_{n,m} \overset{\mathrm{IID}}{\sim} \mathcal{CN}\left(0,1\right)$, $\forall m,n$.
To increase the compressibility of the corresponding channel matrix, we choose a 2$\times$ oversampled DFT grid along the AoA and AoD dimensions, corresponding to resolutions of $\frac{\pi}{N_r}$ and $\frac{\pi}{N_t}$ respectively. We consider a carrier frequency of $f_c=28\mathrm{GHz}$, the maximum CFO limit to be 10ppm of $f_c$ i.e., $f_{\mathrm{max}}=280\mathrm{KHz}$ and the spectral leakage factor($\gamma$) to be $2$. The symbol duration $T$ is chosen to be $0.5\mu s$, which is much larger than the maximum delay spread given in \cite{measrapp} at $28$GHz and hence our narrowband assumption is justified.\\
We assume a digital phase control of 3 bits ($q=8$) at the BS and MS. For each measurement $n\in \left[M\right]$, the vectors $\mathbf{w}_{n}$, $\mathbf{f}_{n}$, have unit $\ell_2$ norm and are chosen independently and uniformly at random from a quantized set. Element wise phase quantization of $3$ bits is performed on the unquantized beamforming weights, also of unit norm. The achievable rate $R=\mathrm{log}_{_2}\left(1+\frac{\rho\left|\mathbf{w}_{\mathrm{est}}^{\ast}\mathbf{H}\mathbf{f}_{\mathrm{est}}\right|^{2}}{\sigma^{2}}\right)$ was averaged over $1000$ realizations of $\mathbf{H}$ for all the three algorithms.
\par Agile Link was evaluated with the same system model and beamtraining vectors as proposed in \cite{agile}. In addition, we perform $3$ bit quantization on these vectors and normalize them. The number of CS measurements required for Agile Link\cite{agile} is given by $B_rB_tN_\mathrm{\mathrm{hash}}$, with $B_r,B_t=O\left(K\right)$ and $N_{\mathrm{hash}}=O\left(\mathrm{log}\left(N_tN_r\right)\right)$. Under the same settings, we also evaluate the standard OMP \cite{OMP} when phase noise and CFO are ignored.
\begin{figure}[h!]
\begin{center}
\includegraphics[width=1 \linewidth]{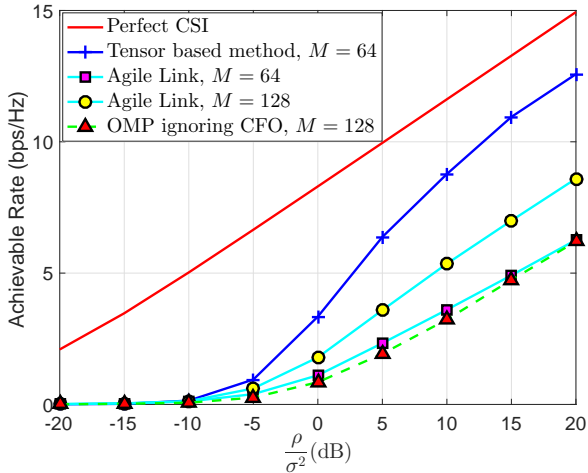}
  \caption{The figure shows the achievable rate versus SNR for our tensor based method, OMP (CFO ignored) and Agile Link for $f_e=265.625\mathrm{KHz}$, $\tau=0.27 \mathrm{rad}$}
  \label{fig:comparison}
\end{center}
\end{figure} 
We evaluate our algorithm in the worst case scenario i.e, when $f_e$ is close to $f_{\mathrm{max}}=280\mathrm{KHz}$ and is maximally off-grid (considering a DFT grid of resolution $\frac{1}{MT}=31.25\mathrm{KHz}$, for $M=64$). Hence $f_e=265.625\mathrm{KHz}$ is chosen.  For Agile Link $B_r, B_t, N_{\mathrm{hash}}$ were optimized and set to $\left(4,4,4\right)$ and $\left(4,4,8\right)$ for $64$ and $128$ measurements respectively. The proposed tensor based approach, however, demands $2P+1$ times higher complexity in memory and time than \cite{agile} or standard OMP, which is well justified by the significant performance gains relative to the other two as seen in Fig.\ref{fig:comparison}. Agile Link considers the magnitude of the noisy measurements which deteriorates its performance at low SNRs. It may be noted that although Agile Link is robust to CFO errors at high SNRs, it needs larger number of measurements to identify the optimal beamforming weights compared to our tensor based approach, because the former ignores the phase of the measurements. Furthermore, our proposed method estimates the channel matrix unlike Agile Link which finds just the beam steering vectors. 
\begin{figure}[h!]
\begin{center}
\includegraphics[width=1 \linewidth]{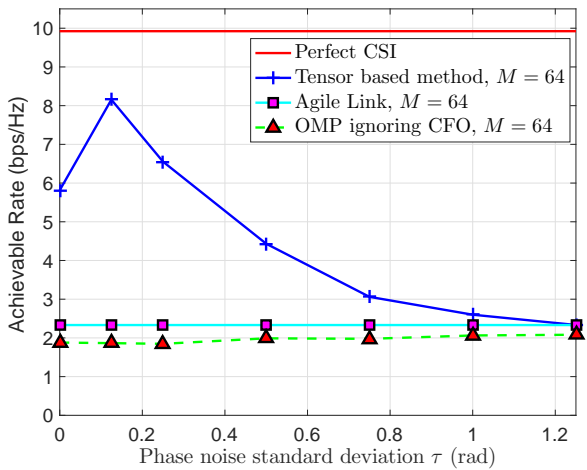}
  \caption{The figure shows the achievable rate versus the standard deviation of phase noise process ($\tau$), for all the three methods, at an SNR of $5$dB}
  \label{fig:tau}
\end{center}
\end{figure} 
From Fig.\ref{fig:tau}, it can be seen that the tensor based algorithm is better than Agile Link in a wide range of $\tau$. The practical value of $\tau$ is given by $2\pi f_{c}\sqrt{cT_{s}}$ \cite{oscill}, where $c=4.7\times 10^{-18}\left(\mathrm{rad.Hz}\right)^{-1}$, and it can be verified that $\tau=0.27\mathrm{rad}$ for our settings. 

\section{Conclusion and future work}
We have proposed a compressive channel estimation technique for narrowband mmWave systems using analog beamforming, that is robust to synchronization impairments. The essence of our paper is to model CFO and channel in a tensor, and to recover the tensor, while still exploiting the sparsity of the mmWave channel. With few measurements, our method is able to do beam alignment better than the existing ones, in addition to estimating the carrier frequency offset. Considering timing mismatch, investigating performance bounds for the proposed method as a function of synchronization impairments, reducing the computational complexity using structured CS algorithms and developing robust CS algorithms for wideband mmWave systems are interesting directions for future work.  
\bibliographystyle{IEEEtran}
\bibliography{refs}

\end{document}